\documentclass[aps,prb,twocolumn,superscriptaddress,nobibnotes,amsmath,amssymb,reprint]{revtex4-2}

\usepackage{graphicx}
\usepackage{xcolor}
\usepackage[normalem]{ulem}
\usepackage{natbib}
\usepackage[english]{babel}

\begin{document}
\title{Low temperature magneto-transport and magnetic properties of MnSb$_2$Te$_4$ single crystals.}
\author{V.N.~Zverev}
\affiliation{Institute of Solid State Physics, RAS, Chernogolovka, 142432, Russia}
\email[e-mail:]{zverev@issp.ac.ru}

\author{N.A.~Abdullaev}
\affiliation{Institute of Physics, Ministry of Science and Education, Baku, AZ1073, Azerbaijan} 
\affiliation{Baku State University, Baku, AZ1148, Azerbaijan}

\author{Z.S.~Aliev}
\affiliation{Institute of Physics, Ministry of Science and Education, Baku, AZ1073, Azerbaijan} 
\affiliation{Baku State University, Baku, AZ1148, Azerbaijan}

\author{I.R.~Amiraslanov}
\affiliation{Institute of Physics, Ministry of Science and Education, Baku, AZ1073, Azerbaijan} 
\affiliation{Baku State University, Baku, AZ1148, Azerbaijan}

\author{Z.A.~Jahangirli}
\affiliation{Institute of Physics, Ministry of Science and Education, Baku, AZ1073, Azerbaijan} 
\affiliation{Baku State University, Baku, AZ1148, Azerbaijan}

\author{I.I~Klimovskikh}
\affiliation{Saint Petersburg State University, 198504 Saint Petersburg} 
\affiliation{Donostia International Physics Center (DIPC), Donostia-San Sebastia?n 20018, Spain}

\author{A.A~Rybkina}
\affiliation{Saint Petersburg State University, 198504 Saint Petersburg} 

\author{A.M.~A.M,Shikin}
\affiliation{Saint Petersburg State University, 198504 Saint Petersburg}  

\author{N.T.~Mamedov}
\affiliation{Institute of Physics, Ministry of Science and Education, Baku, AZ1073, Azerbaijan} 
\affiliation{Baku State University, Baku, AZ1148, Azerbaijan}

Zh.Exp.Teor.Fiz. Vol. 169 (01) (2026) there.

\begin{abstract}
The results of a comprehensive study of MnSb$_2$Te$_4$ single crystals are presented. The structure, Raman spectra, low-temperature transport, Hall effect, magnetization, and magnetic susceptibility are studied. It was established that the crystals are ferromagnetic, with a Curie temperature ranging from 22 to 45\,K for different samples. Hall and magnetization measurements demonstrated that the system is a soft ferromagnet, which is of interest for practical applications.
\end{abstract}

\maketitle

{\bf 1. Introduction}

Magnetic topological insulators have attracted considerable attention from researchers in recent years due to their unusual properties: the possibility of observing the anomalous quantum Hall effect, the topological magneto-electric effect, and dissipationless electron transport in these systems. Another interesting perspective is their use as the basis for spintronics in edge channels and topological qubits. After the discovery \cite{1,2,3,4} of the new magnetic topological insulator MnBi$_2$Te$_4$ a large family of isostructural MnBi$_{2-x}$Sb$_x$Te$_4$ crystals with $0<x<2$ was studied \cite{5,6,7,8,9,10,11,12,13,14,15,16,17,18,19}. For MnSb$_2$Te$_4$ crystals, as for all members of the  MnBi$_{2-x}$Sb$_x$Te$_4$ family with 0<x<2, the term “topological insulators” is used to indicate that topologically protected surface states exist on the surface of these crystals, and that the electronic spectrum contains a Dirac cone and an energy gap, which has been confirmed by numerous calculations. However, there are also calculations that show the possibility of other phases, such as the trivial phase \cite{20}, as well as Weyl I and II type semimetal phases \cite{21}. 

It has been established that crystals of MnBi$_{2-x}$Sb$_x$Te$_4$ with $x=0$ are A-type antiferromagnets (AFM) \cite{1} with a Néel temperature  $T_N \approx $24.5\,K. Their energy spectrum has a gap of about 70\,meV, with the Fermi level lying in the conduction band.   When Bi is completely replaced by Sb, i.e., when $x=2$, there is a significant change in the magnetic and transport properties of these crystals. First of all, an increase in Sb content leads to a crossover from n-type to p-type conductivity due to the Fermi level shifting from the conduction to the valence band. According to \cite{5,6}, the transition from electron to hole conductivity occurs at x $\sim $ 0.6. As the Sb content increases, the magnetic properties of these crystals also evolve: MnBi$_2$Te$_4$ samples are antiferromagnetic, while for MnSb$_2$Te$_4$ samples, according to the literature, there are data on the existence of both AFM and ferromagnetic (FM) states \cite{7,8}. The ideal ground state of this crystal should be antiferromagnetic \cite{9,11}. According to experimental data \cite{6} when Bi was completely replaced by Sb, the crystals remained antiferromagnetic, but the $T_N$ temperature dropped to approximately 19\,K. However, as shown in \cite{12,13}, MnSb$_2$Te$_4$ crystals can become ferromagnetic due to the presence of Mn/Sb substitution defects. Indeed, ferromagnetism in MnSb$_2$Te$_4$ samples has now been confirmed by neutron diffraction experiments, magnetic measurements, and direct observation of magnetic domains \cite{12,14,15,17}. In \cite{14} a record Curie temperature of approximately 50\,K was obtained on thin-film ferromagnetic MnSb$_2$Te$_4$ samples grown by molecular epitaxy. On single-crystal FM samples with a non-stoichiometric composition, it is possible to achieve even higher values of $T_C \approx $58\,K \cite{18}. In the experimental works mentioned above, the samples were grown using different methods, and their properties were measured using different experimental techniques, which complicates the comparative analysis necessary to understand the reasons for the formation of FM and AFM phases and to control critical temperatures. In this work, we investigated for the first time a series of MnSb$_2$Te$_4$ crystals in a wide range of Curie temperatures (22–45\,K), which allowed us to track the evolution and interrelation of FM and AFM orderings. The results of comprehensive studies of single-crystal MnSb$_2$Te$_4$ samples, selected and characterized using X-ray structural analysis and Raman spectroscopy, are presented. To study the electronic and magnetic properties of crystals, low-temperature transport, the Hall effect, field dependencies of magnetization, and magnetic susceptibility were investigated.

{\bf 2. Experimental results and discussion}

{\bf 2.1. Characterization of samples}

The magnetic, transport and magneto-transport properties of MnSb$_2$Te$_4$ were investigated on samples selected and characterized by X-ray diffraction and Raman scattering methods at room temperature. MnSb$_2$Te$_4$ single crystals were obtained by cleaving ingots grown by directional crystallization. The samples were approximately 2x1x0.1\,mm$^3$ in size. Seven samples were selected from all characterized MnSb$_2$Te$_4$ crystals for subsequent transport and magnetic measurements. The numeration of the samples was retained for all measurements and is indicated in the presentation of the experimental results. 

X-ray diffractometric studies of the obtained samples were performed on a BRUKER XRDD2 Phaser X-ray diffractometer using CuK$_\alpha $ radiation. The phase composition and structural analysis of the samples were performed using the EVA and TOPAS-4.2 software packages (Bruker, Germany). The obtained diffractograms fully corresponded to the calculated ones and confirmed the tetradimite structure R3m \cite{18}. The lattice parameters of the studied crystals were determined with an accuracy of ±0.001\,$\mathring{A}$. Figure 1 shows the X-ray diffractogram of a  MnSb$_2$Te$_4$ single crystal. 

\begin{figure}
\center \includegraphics[width=0.5\textwidth]{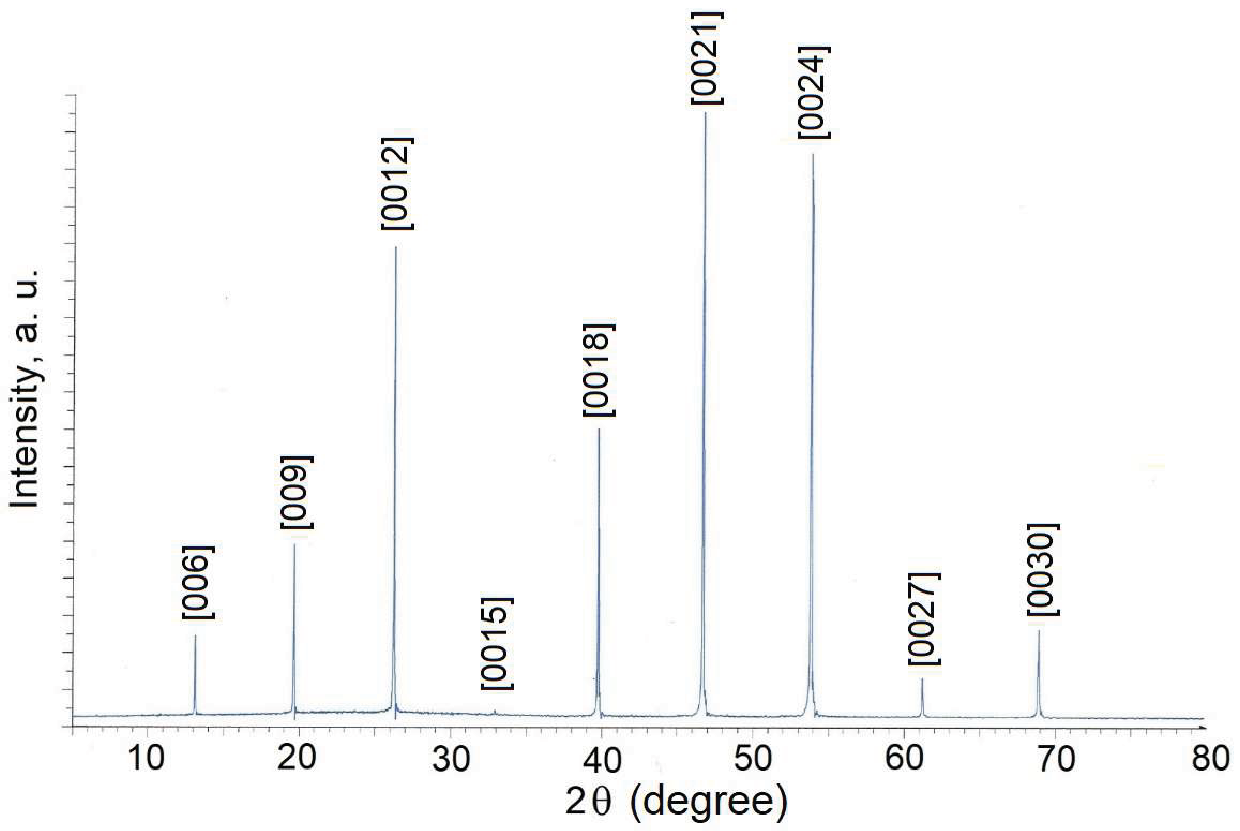}
\caption{Diffractogram of aMnSb$_2$Te$_4$ single crystal.}
\end{figure}

The study of Raman light scattering in MnSb$_2$Te$_4$ crystals was carried out on samples with a mirror surface in reverse scattering geometry using a Nanofinder 30 confocal Raman microspectrometer (Tokyo Instr., Japan). A YAG:Nd laser with a second harmonic emission wavelength of  $\lambda $=532\,nm and a maximum power of 10\,mW was used as the excitation light source. The excitation spot diameter was 4\,$\mu m$ when using a ×100 magnification lens. A charge-coupled device (CCD camera Newton, Andor Technology) cooled thermoelectrically to a temperature of -100\,°C was used as a radiation detector. The signal accumulation time was typically about one minute. The samples did not degrade at the radiation power used (6-8\,mW). 

\begin{figure}
\center \includegraphics[width=0.5\textwidth]{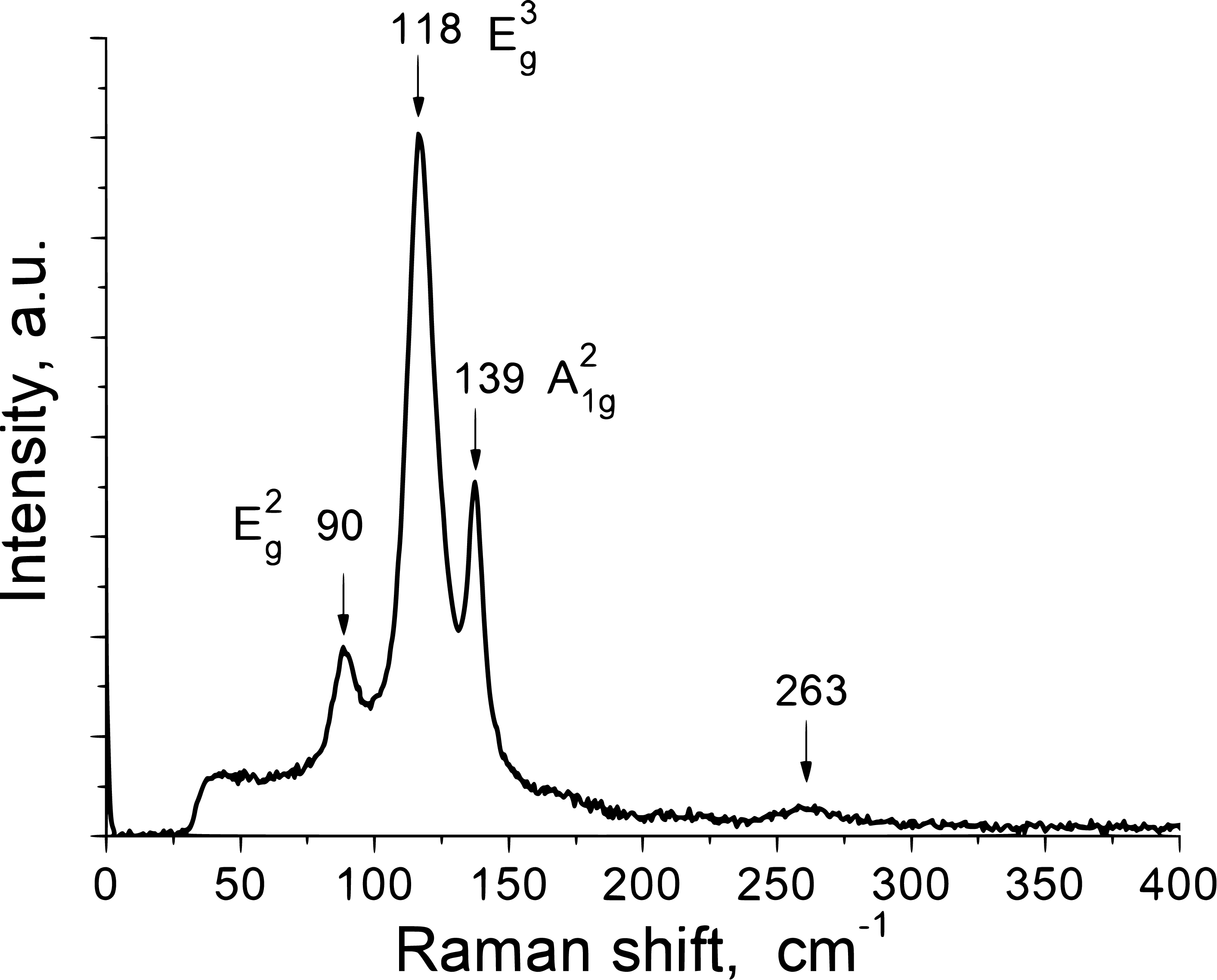}
\caption{Raman scattering spectrum in MnSb$_2$Te$_4$ single crystals}
\end{figure}

\begin{figure}
	\center \includegraphics[width=0.5\textwidth]{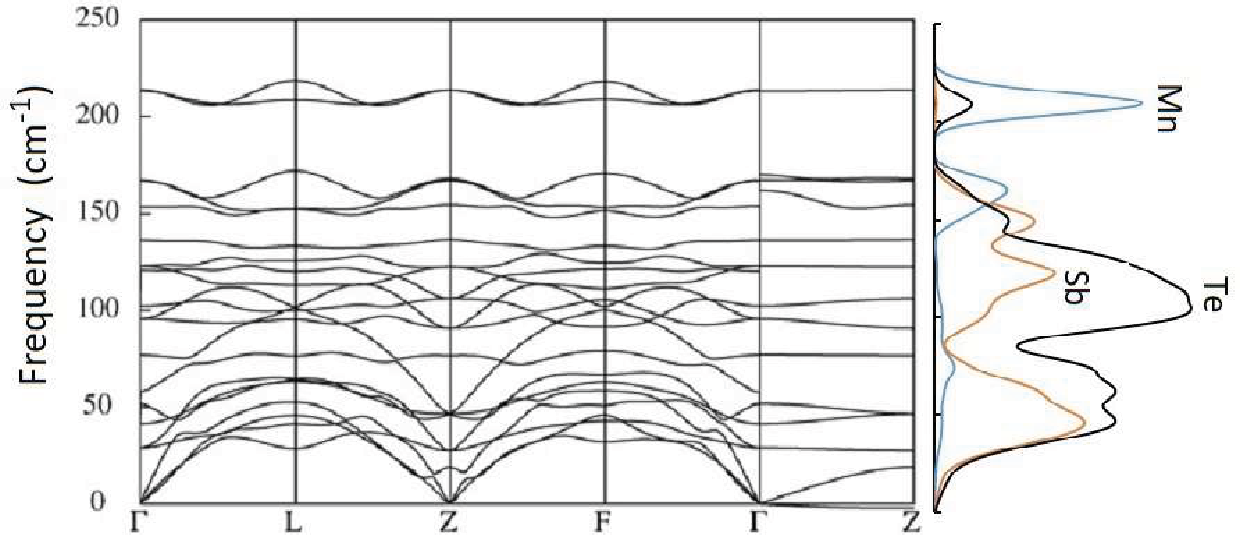}
	\caption{(color online) Dispersion of phonon modes in MnSb$_2$Te$_4$ crystals (left) and partial density of states (PDOS) projected onto Mn, Sb, and Te atoms (right).}
\end{figure}

We performed theoretical calculations based on the first principles of the dynamics of the crystal lattice of MnSb$_2$Te$_4$ crystals. The calculations were performed within the framework of the density functional perturbation theory \cite{19}. Fig. 3 shows the dispersion of phonon modes in MnSb$_2$Te$_4$ crystals and the partial density of phonon states (PDOS) projected onto Mn, Sb, and Te atoms.  

 \begin{figure*}
	\center \includegraphics[width=1\textwidth]{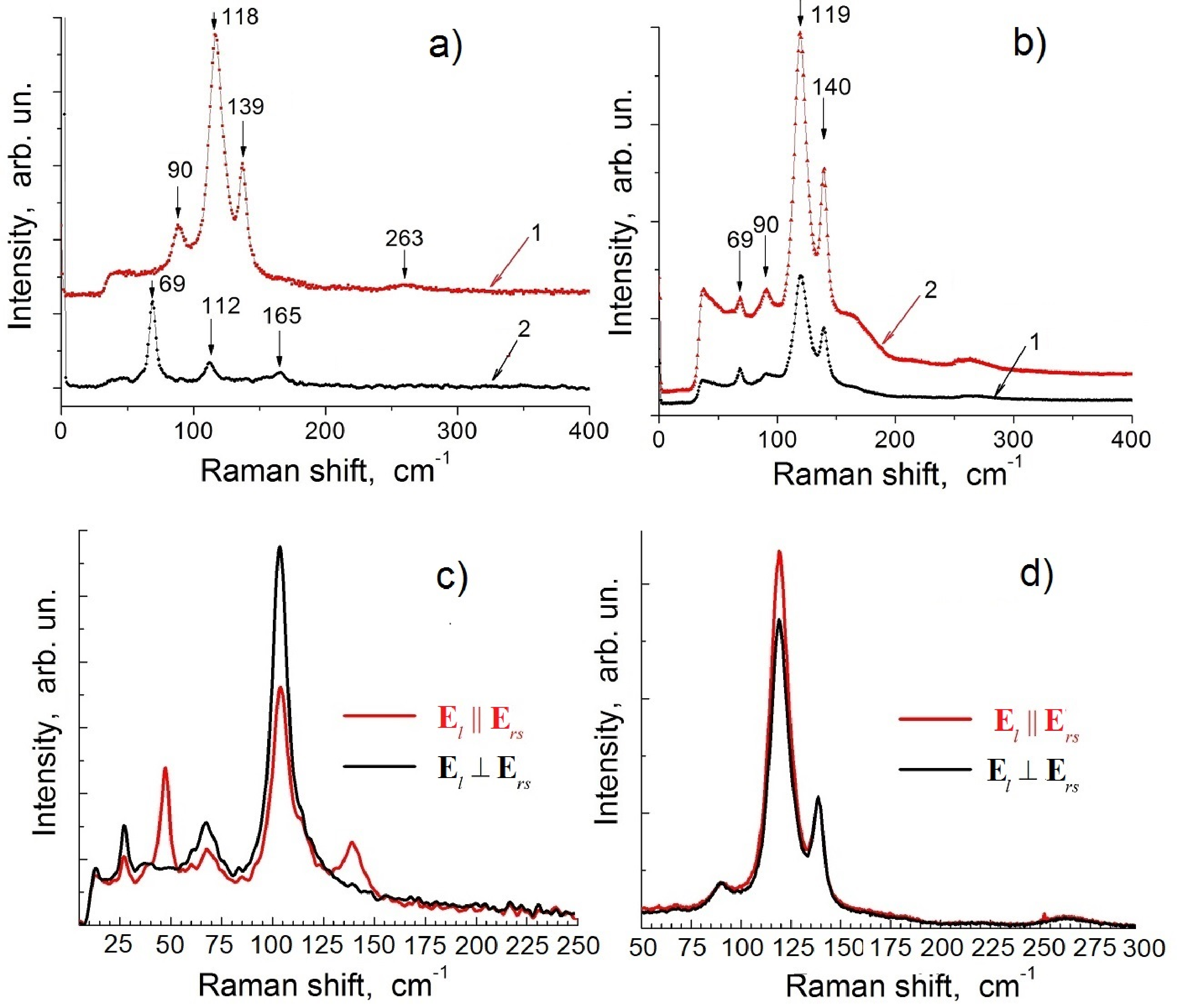}
	\caption{(color online) Raman scattering spectra of crystals: a)  MnSb$_2$Te$_4$ (1) and Sb$_2$Te$_3$ (2); b) Sb$_2$Te$_3$ in two different defect regions (1 and 2), c) MnBi$_2$Te$_4$ at different polarization of the exciting laser radiation ($\bf {E}_l$ and $\bf {E}_{rs}$ are the exciting laser field in the plane of the sample surface and the scattered light field, respectively); d) MnSb$_2$Te$_4$ at different polarization of the exciting laser radiation.} 
\end{figure*}

As can be seen in Fig. 3, at high frequencies (~200\,cm$^{-1}$), lighter Mn atoms make the main contribution to the phonon density of states, while at low and medium frequencies (up to\,150 cm$^{-1}$), Sb and Te atoms make the predominant contribution.

Theoretical calculations have shown the presence of six Raman-active modes in MnSb$_2$Te$_4$ crystals, as in MnBi$_2$Te$_4$ \cite{22}. These are frequencies of 29, 94, 123\,cm$^-1$ and 52, 136, 167\,cm$^-1$ for the E$_g$ and A$_{1g}$ modes, respectively. However, unlike MnBi$_2$Te$_4$, only two E$_g$ modes, 90\,cm$^-1$ (94) and 118\,cm$^-1$ (123), and only one A$_{1g}$ mode, 139\,cm$^-1$ (136), are experimentally observed in MnSb$_2$Te$_4$  crystals. The consistently observed mode at 263\,cm$^-1$ is far beyond the upper limit of possible frequencies for vibrational modes in MnSb$_2$Te$_4$, as can be clearly seen from the calculated phonon density of states (Fig. 3). 

Unlike the isostructural compound MnBi$_2$Te$_4$, which is an antiferromagnetic topological insulator \cite{1}, MnSb$_2$Te$_4$ compounds exhibit ferromagnetic ordering of the spins of magnetic atoms. As already noted in the Introduction, this is explained by the presence of a large number of structural defects caused by the mutual substitution of Mn and Sb atoms. Our studies of Raman scattering confirm the validity of this assumption. It is worth noting four main differences between combinational light scattering in MnSb$_2$Te$_4$ and  MnBi$_2$Te$_4$. First, since, according to theoretical calculations \cite{22}, Mn atoms are immobile in the shifts of active modes of Raman scattering, the Raman scattering spectra of MnBi$_2$Te$_4$ crystals are very similar to those for Bi$_2$Te$_3$ \cite {19,22}. At the same time, the Raman scattering spectra of MnSb$_2$Te$_4$ \cite {23} differ significantly from those of  Sb$_2$Te$_3$ (Fig. 4(a)). Secondly, in the defective regions of two different Sb$_2$Te$_3$ samples (Fig. 4(b)), the spectra are very similar to those of MnSb$_2$Te$_4$ crystals (Fig. 1(a)), while the low-frequency mode of 69\,cm$^{-1}$ Sb$_2$Te$_3$ is apparently preserved, i.e., there are areas in the MnSb$_2$Te$_4$ crystal where Mn is replaced by Sb. Thirdly, unlike the spectra of MnBi$_2$Te$_4$ (Fig. 4(c)), no polarization dependencies are observed in the spectra of MnSb$_2$Te$_4$ \cite{23} (Fig. 4(d)), which indicates a significantly higher density of defects  in MnSb$_2$Te$_4$ samples. And fourthly, the theoretically calculated frequencies of modes for ideal crystals of MnSb$_2$Te$_4$ are only partially observed, and the mode with a frequency of 263\,cm$^{-1}$ is far beyond the upper limit of possible frequencies for vibrational modes in MnSb$_2$Te$_4$. Thus, the listed data on combinational scattering are consistent with the previously obtained results of other authors on the presence of a large number of structural defects caused by the mutual substitution of Mn and Sb atoms. 

{\bf 2.2. Transport and magnetotransport  properties.}

Contacts for transport measurements were prepared using conductive graphite paste. The samples were mounted in an insert allowing operation over a wide temperature range (1.4-300)\,K, immersed in a liquid helium cryostat with a superconducting solenoid. The field in the magnetotransport and magnetic measurements was oriented perpendicular to the sample plane, i.e., along the c-axis of the crystal. Resistivity and Hall effect measurements were performed using a Lock-in detector according to the standard four-probe technique in Hall bridge geometry at an alternating current frequency of 20\,Hz, with a measuring current not exceeding 1\,mA. 

\begin{figure}
	\center \includegraphics[width=0.5\textwidth]{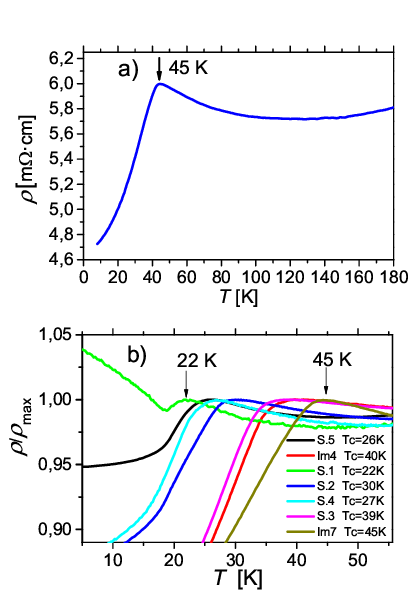}
	\caption{(color online) a) Temperature dependence of the resistivity of MnSb$_2$Te$_4$ sample with $T_C=45\,$K. (b) Normalized resistivity $\rho/\rho_{max}$ of samples obtained in different syntheses, $\rho_{max}$ - resistivity of samples in the local maximum at $T=T_C$.}
\end{figure}

\begin{figure}
	\center \includegraphics[width=0.5\textwidth]{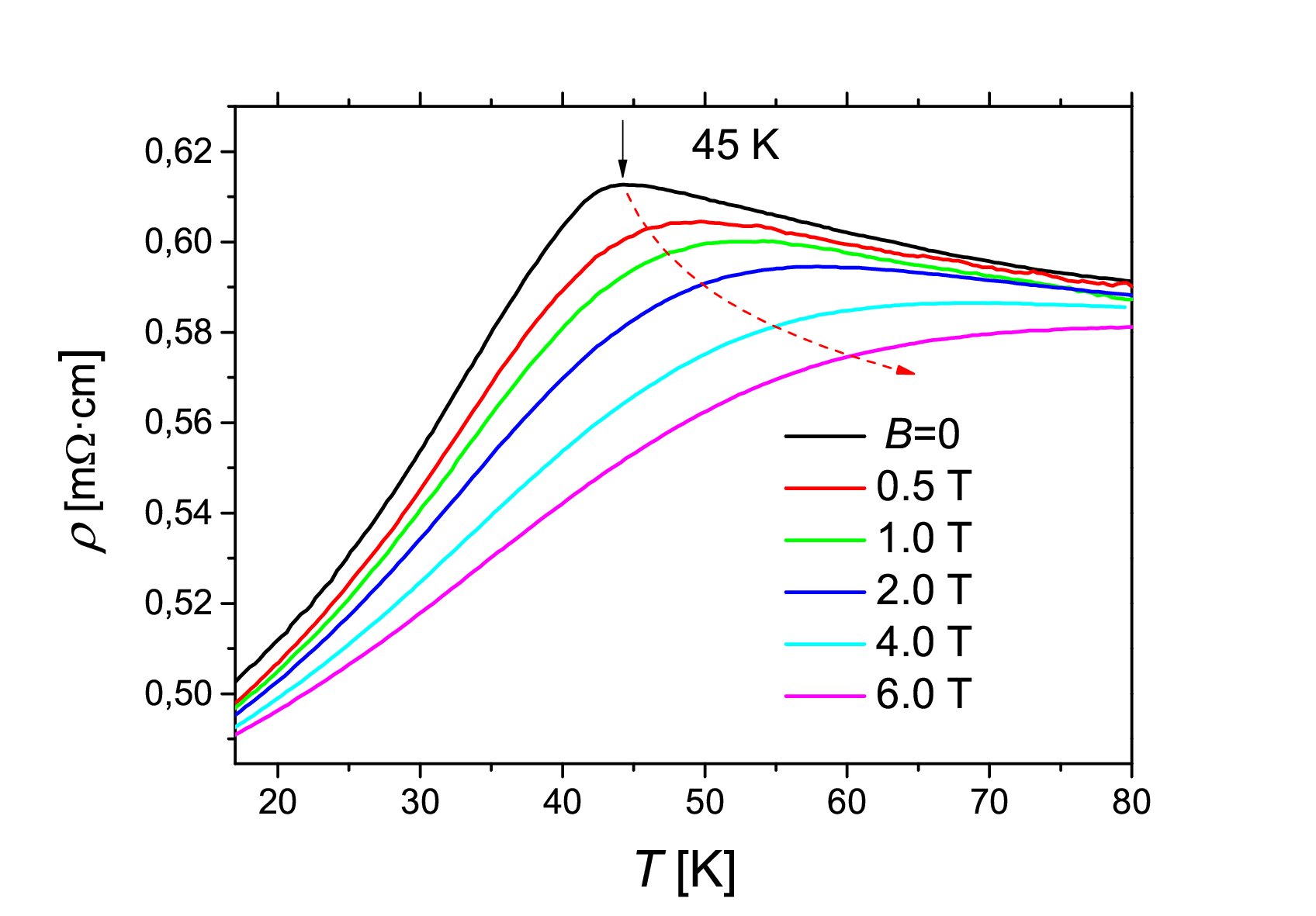}
	\caption{(color online) Effect of the magnetic field on Curie temperature. The shift in the position of the maximum in the dependence $\rho(T)$ is shown by the red dotted arrow.}
\end{figure}

The MnSb$_2$Te$_4$ samples grown and studied in this work are, as shown by all the results below, ferromagnets with a Curie temperature $T_C$ that varies widely depending on the quality of the sample. The temperature dependence of the resistance of MnSb$_2$Te$_4$ samples shows a maximum caused by the amplification of magnetic fluctuations in the FM transition region (Fig. 5(a)). Fig. 5(b) shows the dependencies of the normalized resistance $\rho /\rho _{max}$ obtained on different MnSb$_2$Te$_4$ samples ($\rho _{max}$ is the resistivity of the samples at the maximum at $T=T_C$). As can be seen from the figure, the Curie temperature $T_C$, indicated by arrows, varies among samples obtained in different syntheses, with the maximum value of $T_C$ reaching 45\,K, which, to our knowledge, is a record value for bulk single crystals of stoichiometric composition. It should be noted that this value of $T_C$ in MnSb$_2$Te$_4$ significantly exceeds the Néel AFM transition temperature $T_N\approx $24.5\,K in MnBi$_2$Te$_4$ \cite {1}. In Fig. 5(b), curves with $T_C \le 30$\,K also show a break in the 19\,K temperature range, which intensifies with decreasing $T_C$. We attribute this to the presence of an AFM phase in the samples, which will be discussed in more detail in the section {\bf 2.3.}

 \begin{figure}
	\center \includegraphics[width=0.5\textwidth]{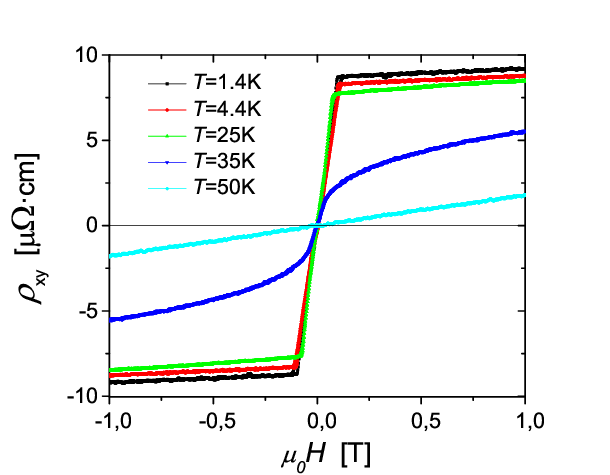}
	\caption{(color online) Dependence of the Hall resistance $\rho_{xy}$ on the magnetic field at different temperatures for the sample Im4 with $T_C$= 40\,K.}
\end{figure}

All MnSb$_2$Te$_4$ samples studied exhibited p-type conductivity, with the concentration of charge carriers measured by the Hall effect at low temperatures ranging (1-5)$\cdot$10$^{19}$\,cm$^{-3}$. The Hall concentration was determined from the slope of the linear dependence of $\rho _{xy}$ on the magnetic field at $T$, slightly exceeding $T_C$. The mobility in most of the samples studied ranged from several tens to hundreds of cm$^2$/(V$\cdot$s).

As mentioned above, in MnSb$_2$Te$_4$ crystals Mn/Sb substitution defects are the main source of disorder affecting $T_C$. Additional confirmation of the relationship between sample defectivity and $T_C$ is provided by the saturation of the temperature dependence and even an increase in resistance with decreasing temperature at $T<T_C$ in some samples with low critical temperatures (see curves for samples S.1 and S.5 in Fig. 5(b)). The negative derivative on the dependence $\rho (T)$ for sample S.1 may be related to the weak localization effect caused by quantum corrections to conductivity. However, for samples with a reduced $T_C$ value, according to \cite{24}, the appearance of a section with a negative derivative on the dependence $\rho (T)$ may also be due to the appearance of a fraction of the AFM phase, in which the Fermi level falls into the forbidden zone, which affects the resistance. 

In a magnetic field, the maximum in the temperature dependence of resistance widens, and this feature shifts to higher temperatures as the field increases (Fig. 6).  This behavior in a magnetic field is consistent with our observation \cite{16} of a similar phenomenon in ferromagnetic crystals of the system (MnBi$_2$Te$_4$)(Bi$_2$Te$_3$)$_m$ at m$>$3 and has a clear physical reason: uniaxial magnetic anisotropy and an external magnetic field act together, stabilizing the FM state at a higher temperature. 

Important information about the magnetic properties of the samples under study, along with magnetic measurements (see Section 2.3), is provided by measurements of Hall conductivity $\rho_{xy}$, since the Hall voltage signal is determined by the sum of the external applied field and the internal field created by paramagnetic Mn atoms.  Fig. 7 shows the $\rho_{xy}(H)$ curves recorded at different temperatures both below and above $T_C$ on a sample with a value of $T_C$=40\,K. As can be seen from the figure, at $T<T_C$ $\rho_{xy}(H)$ exhibits dependencies characteristic of ferromagnets, reflecting the behavior of the sample magnetization. At the same time, the curves recorded when the direction of the magnetic field sweep is changed practically coincide, i.e., the hysteresis loop on this sample is very narrow, its width does not exceed 50\,Oe. Hysteresis loops on samples with lower $T_C$ values were also quite narrow, with a width not exceeding 200\,Oe. Thus, it can be stated that the MnSb$_2$Te$_4$ crystal is a distinctly soft ferromagnetic material. It should be noted, however, that the value of the hysteresis loop width measured in a setup with a superconducting solenoid may contain an error associated with the magnetic flux frozen in the solenoid.

{\bf 2.3. Magnetic properties.}

The field and temperature dependences of the magnetization of MnSb$_2$Te$_4$ samples were measured on an SQUEED magnetometer MPMS (Quantum Design). The external magnetic field in magnetic measurements was oriented perpendicular to the plane of the sample, i.e., along the $c$-axis of the crystal. Magnetic susceptibility measurements were performed in zero-field-cooled (ZFC) mode (with precooling of the sample to 2\,K in the absence of a field and subsequent measurement of magnetization in a field during heating from 2 to 300\, K) and field-cooled (FC) (measurement of magnetization during cooling from 300 to 2\,K in the same applied field). Measurements of temperature dependencies of magnetization were carried out in weak (0.005 and 0.1\,T) and strong (5\,T) magnetic fields to obtain comprehensive information about the magnetic properties of the crystal, especially if it contain mixed magnetic phases. Four samples of MnSb$_2$Te$_4$ numbered S.1–S.4 were studied (see Fig. 5 b). Therefore, for comparison of data for different samples, Fig. 8 shows graphs of magnetic susceptibility as a function of temperature, measured in a magnetic field of 0.005\,T, and their representation as the first derivative with respect to temperature to demonstrate critical temperatures. 

 \begin{figure*}
\center \includegraphics[width=1\textwidth]{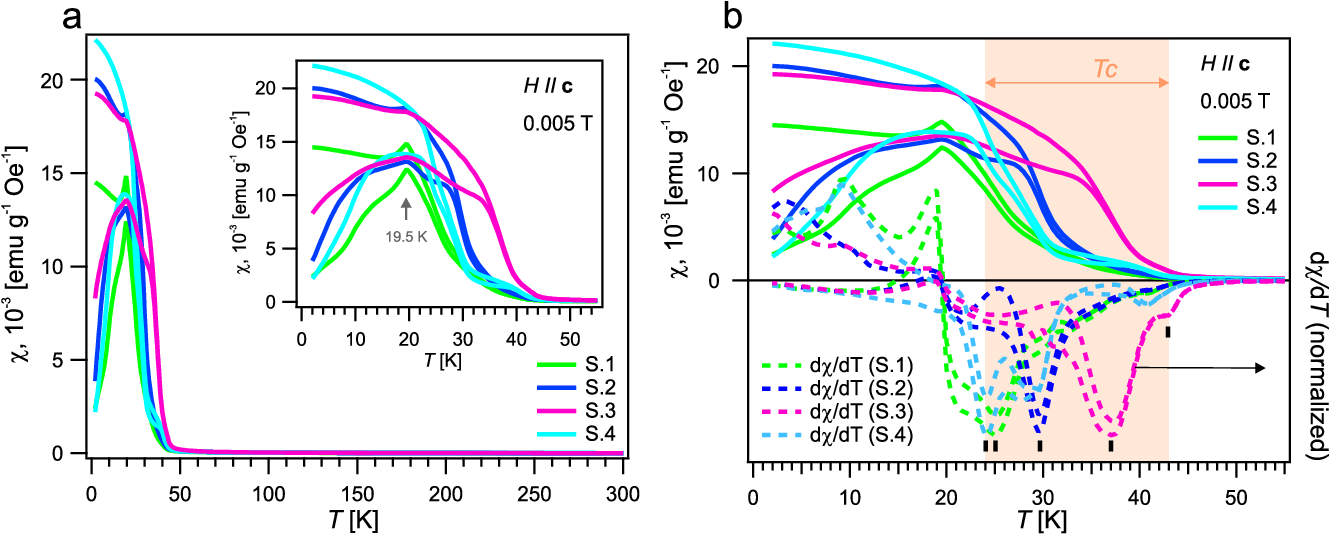}
\caption{(color online) Temperature dependence of magnetic susceptibility $\chi (T)$ (a) and its first derivative with respect to temperature for a series of MnSb$_2$Te$_4$ samples (b). The inset in Fig.8a shows the dependence $\chi (T)$ on an expanded scale in the low-temperature region. In Fig.8b, intensity-normalized graphs of $d\chi /dT(T)$ (right scale) are shown simultaneously with $\chi (T)$ for clarity.}
\end{figure*}

As can be seen in Fig. 8, the series of measured samples is characterized by different Curie temperatures $T_C$, similar to the results of magnetotransport measurements (see Fig. 5(b)). At the same time, sample S.3 has the highest critical temperature $T_C$ among the samples studied, reaching $\sim$ 40\,K. Note that the Curie temperature values obtained from transport and magnetic measurements differ slightly, since the procedures for determining $T_C$ differ.

 In addition to FM ordering, signs of AFM one can also be detected in all samples. This is evidenced by the peak at 19.5\,K on the magnetic susceptibility curves (Fig. 8(a)). This peak, corresponding to the Néel temperature, has been previously observed in the literature for MnSb$_2$Te$_4$ samples, depending on the crystal growth conditions \cite {9, 10, 11, 12}. Also, earlier in \cite {8}, it was shown that in MnBi$_{2-x}$Sb$_x$Te$_4$ samples, when Bi atoms are replaced by Sb atoms, a mixed magnetic structure is observed, i.e., there is a phase with both FM and AFM ordering, revealed by SQUID magnetometry and ferromagnetic resonance methods. It should be noted that, as can be seen in Fig. 8, the Curie temperature is higher when the AFM contribution is lower (peak at 19.5\,K). This is consistent with the transport data (Fig. 5(a)) and corresponds to a higher concentration of substitution defects. 
 
Fig. 9 shows the field dependencies of magnetization $M(H)$ for a series of samples studied, measured at a temperature of 1.8\,K. It can be seen that the shape of the curves corresponds to the results of magnetotransport measurements and represents a very narrow hysteresis loop (the width of the loop will be analyzed in detail below). The contribution of the AFM phase is relatively small for these samples. Only for sample S.1 is there a noticeable deviation in the shape of the magnetization curve and hysteresis loop and the presence of breaks characteristic of the AFM phase. This behavior may indicate fewer substitution defects in this sample and, as a result, a greater contribution of AFM ordering in the crystal. At the same time, on the magnetic susceptibility curves, the most distinct AFM peak of the $T_N$=19.5\,K transition is also observed for sample S.1. In the other samples, the FM phase is predominantly observed. 

\begin{tabular}{|c|c|c|c|c|}
\hline
sample & S.1 & S.2 & S.3 & S.4 \\
\hline
$H_{crc}$ [Oe]) & 209 & 100 & 64 & 205 \\
\hline
$T_N$ [K] & 19.5 & 19.5 &19.5 & - \\
\hline
$T_c$ [K] from $\chi (T) $ & 25, 33 & 29.5 & 37, 43 & 24, 29.5, 40.5 \\
\hline
$T_c$ [K] from $\rho  (T) $ & 22 & 30 & 39 & 27 \\
\hline
\end{tabular}

\begin{figure}
\center \includegraphics[width=0.5\textwidth]{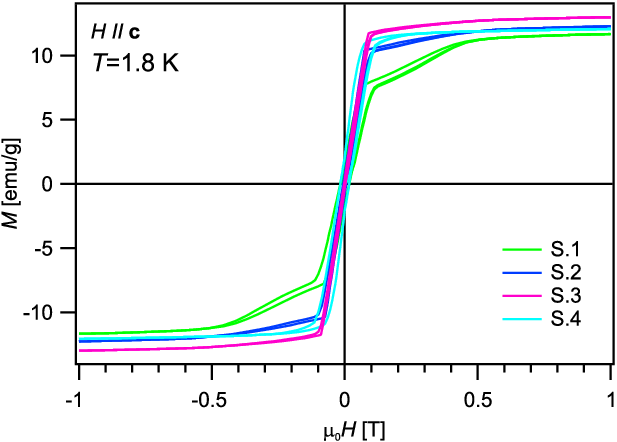}
\caption{(color online) Field dependencies of magnetization $M(H)$ for a series of MnSb$_2$Te$_4$ samples measured at $T$=1.8\,K in a field applied along the crystallographic c-axis.}
\end{figure}

\begin{figure*}
	\center \includegraphics[width=1\textwidth]{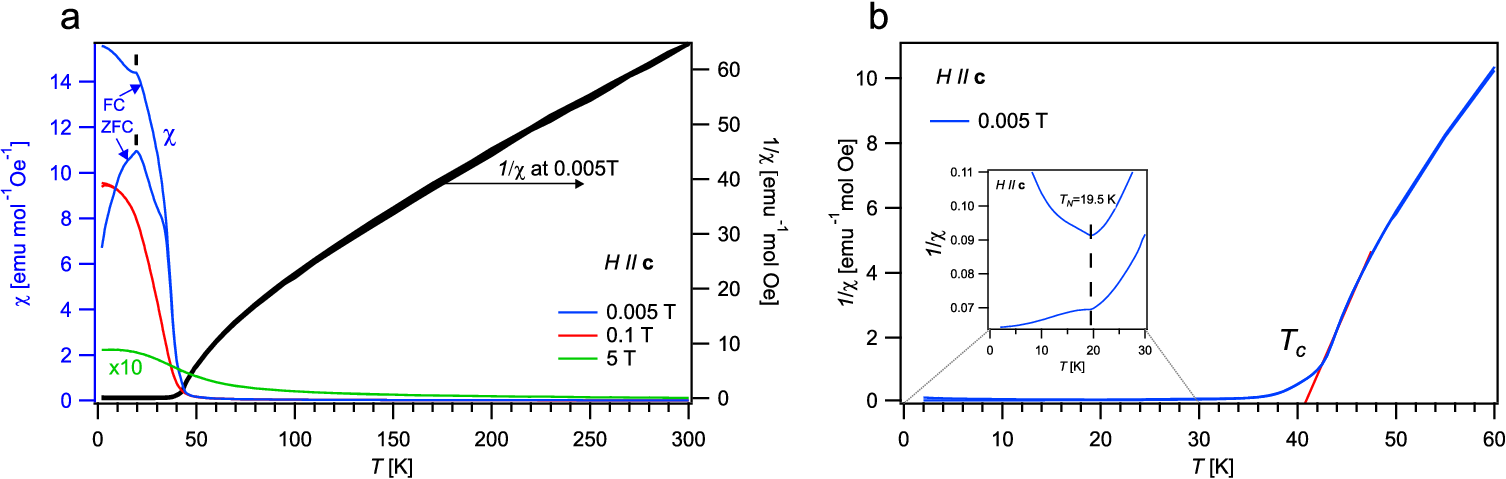}
	\caption{(color online) (a) - Temperature dependence of magnetic susceptibility $\chi (T)$ (left scale) and inverse magnetic susceptibility $1/\chi (T)$ (right scale) for sample S.3, (b) – inverse magnetic susceptibility as a function of temperature, shown on an expanded scale.}
\end{figure*}

For a comparative analysis of critical temperatures obtained from transport and magnetic measurements, we present these values in Table 1 for four measured samples. The determination of the Curie temperature from the $\chi (T)$ dependencies is ambiguous (see below), since some samples exhibit several peculiarities, which may be related to the presence of several phases with different densities of defects. However, for the most prominent peaks (see the first derivative $\chi (T)$ in Fig. 8), there is a good agreement with the critical temperatures determined from magnetotransport measurements. Table 1 also shows the coercive forces for each sample obtained from the $M(H)$ curves. It can be seen that the coercive force is greater for samples with a lower Curie temperature, which may be due to lower defectivity and, accordingly, larger magnetic domains. 

Next, for detailed analysis and comparison with magnetotransport measurements, the FM phase in the studied MnSb$_2$Te$_4$ crystals obtained by different synthesis methods will be analyzed using the example of a sample with a transition temperature $T_C\sim $ 40\,K (sample S.3). Fig. 10 shows the temperature dependencies of magnetic susceptibility $\chi (T)$ and inverse magnetic susceptibility $1/\chi (T)$ measured in different magnetic fields (0.005\,T, 0.1\,T, and 5\,T). The dependence of the inverse magnetic susceptibility in the 0.005\,T field is shown on the same graph on the second axis (in black).

In magnetic fields of 0.1\,T and 5\,T, the zero-field cooling (ZFC) and field cooling (FC) magnetic susceptibility curves coincide across the entire temperature range (Fig. 10), indicating rapid saturation of the magnetic moment with increasing field. However, in the case of a weak field of 0.005\,T, the ZFC and FC curves diverge in the region $T<T_C$. This divergence may indicate the presence of phases in which residual magnetization is observed in the sample, for example, of the spin glass type. The magnetic susceptibility curve $\chi (T)$ increases sharply in the $T$ range below $\sim $ 45\,K, indicating a transition from the paramagnetic to the ferromagnetic phase (Fig. 10(a)). At the same time, at a temperature of 19.5\,K, a maximum is observed on the  $\chi (T)$ graph, indicating a transition to an antiferromagnetic state. The Curie temperature of about 40\,K and the Néel temperature of about 19.5\,K were also determined from the graphs of the inverse magnetic susceptibility  $1/\chi (T)$ (Fig. 10(b)). 

The nonlinearity of the $\chi (T)$ graph at 0.005\,T makes it difficult to approximate it using the Curie-Weiss law. Therefore, to determine the parameters of the Curie-Weiss equation, the temperature dependence $1/\chi (T)$ measured in a 5\,T field (i.e., in the saturation region of the M(H) dependence in Fig. 9) was used. The results of the analysis are presented in Fig. 11. In this case, the high-temperature region of the graph at 5\,T can be approximated by a linear dependence obeying the Curie-Weiss law, 
$$
\chi =\chi_0 + \frac{C}{T-T_{\theta}} ,
$$
where $\chi_ 0$ is a temperature-independent term, C is the Curie constant, and $T_{\theta}$ is the paramagnetic Curie-Weiss temperature. As a result of linear approximation, the following parameters were determined: Curie constant C = 3.73 emu$\cdot$K/mol$\cdot$Oe and paramagnetic Curie temperature $T_{\theta}$=8.8\,K. The Curie-Weiss temperature turned out to be positive, which indicates a predominantly ferromagnetic ordering of moments in high fields. At the same time, approximation according to the Curie-Weiss law of the dependence $1/\chi (T)$ in a weak field (0.005\,T) in the high temperature range (200-300\,K) gives a negative value of the Curie-Weiss temperature, which characterizes the antiferromagnetic ordering of magnetic moments in weak fields. 

\begin{figure*}
\center \includegraphics[width=1\textwidth]{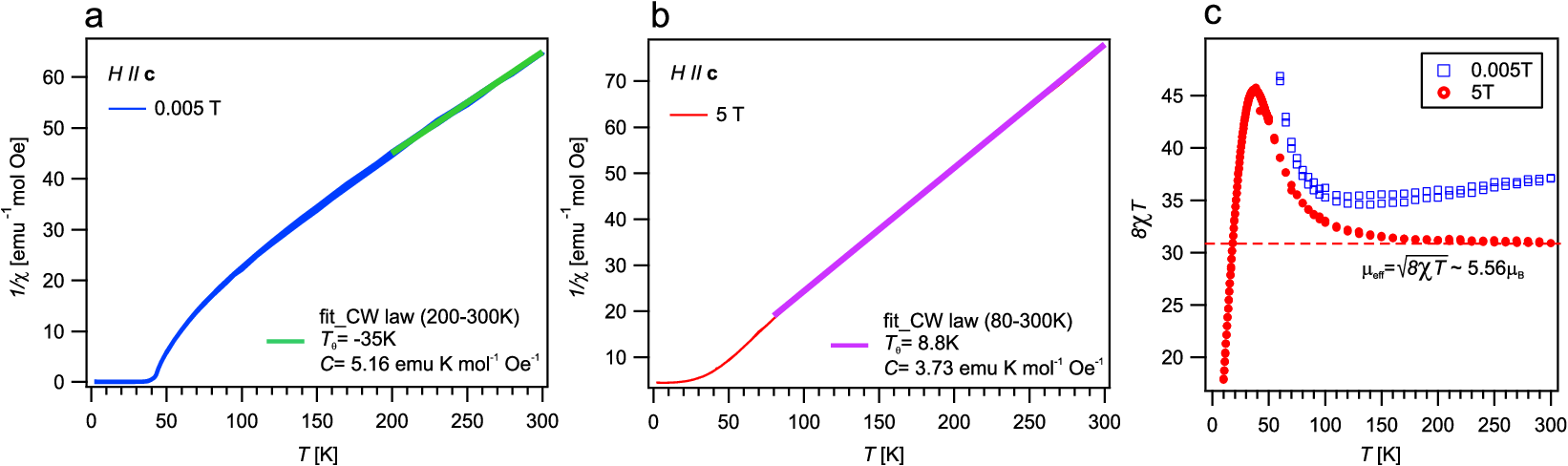}
\caption{(color online) Approximation of the inverse magnetic susceptibility graph by the Curie-Weiss linear law for the dependence $1/\chi (T)$, measured at a) 5\,T and b) 0.005\,T; c) determination of the effective magnetic moment per magnetic center.}
\end{figure*}

The effective magnetic moment $\mu_{eff}$ per paramagnetic center (the center at which the magnetic moments are localized) was determined. Two methods were used to determine $\mu_{eff}$: 1) based on the found value of the Curie constant using the formula $\mu ^2$=8$C$; 2) based on the graph $\mu ^2$=8$\chi T$ (shown in Fig. 11(c)). The following values of the effective moment were obtained:

1)	$\mu _{eff}$= 5.46$\mu B$/Mn

2)	$\mu _{eff}$= 5.56$\mu B$/Mn

\begin{figure*}
\center \includegraphics[width=1\textwidth]{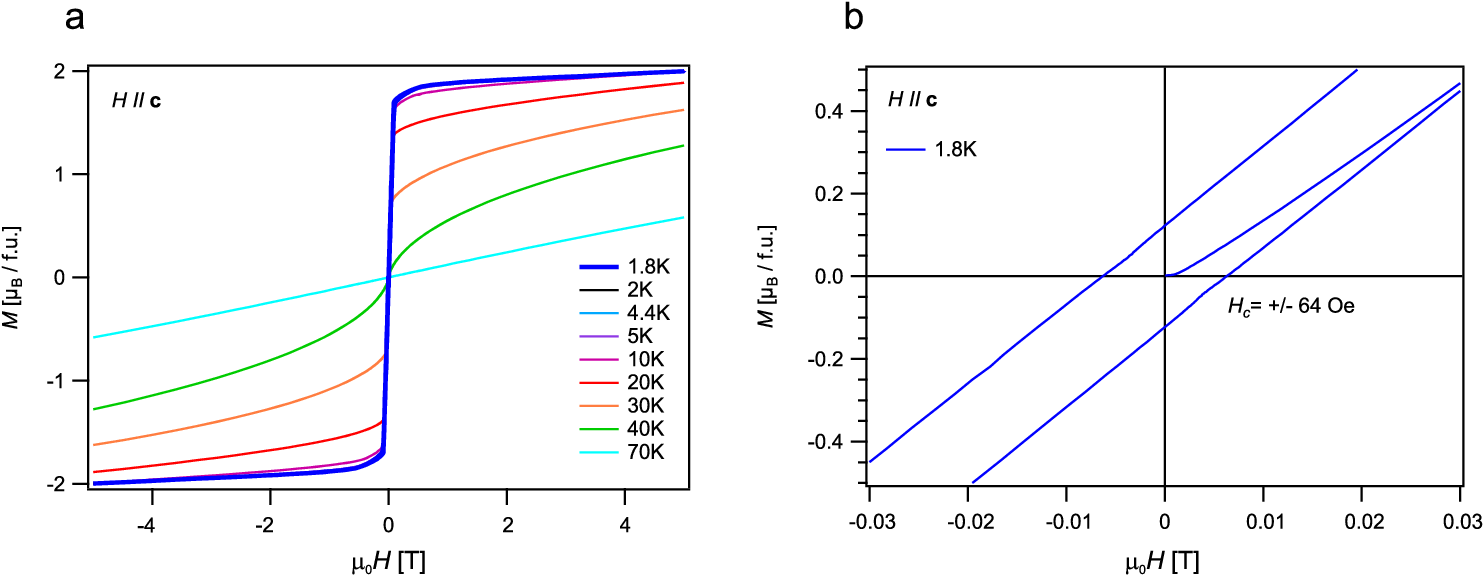}
\caption{(color online) (a) Field dependence of magnetization $M(H)$ for MnSb$_2$Te$_4$ (sample S.3) at different temperatures and (b) at $T$ = 1.8\,K on an expanded scale to demonstrate the hysteresis loop.}
\end{figure*}

Fig. 12 shows the magnetization isotherms $M(H)$ measured for sample S.3 in a magnetic field applied along the $c$ axis. At $T < T_C$, the shape of the $M(H)$ curves characteristic of ferromagnetism is observed. At low temperatures, narrow hysteresis loops are observed (coercive force 64\,Oe at 1.8\,K after normalization to the instrumental curve). At temperatures $T > T_C$, a nonlinear dependence is observed. The value of the magnetic moment at saturation is $2\mu B$/form.unit at 5\,T, which is consistent with previously observed values for MnSb$_2$Te$_4$ crystals \cite {12, 14}. 

{\bf Conclusion}

In conclusion, we note that this work presents comprehensive studies of single-crystal MnSb$_2$Te$_4$ samples, including investigations of Raman scattering spectra, low-temperature transport, magnetotransport, magnetization, and magnetic susceptibility. It has been shown that all the samples studied are ferromagnets, while our data also demonstrate a small contribution of antiferromagnetic behavior, which is due to the presence of phases with different concentrations of Mn/Sb substitution defects. The presence of Mn/Sb defects is consistent with the results previously obtained by other authors, as well as with the analysis of the Raman scattering spectra we studied. The Curie temperature ($T_C$) varies among samples obtained in different syntheses, with the maximum $T_C$ reaching 45\,K, which, to our knowledge, is a record value for bulk single crystals of stoichiometric composition.  The results of Hall effect and magnetization measurements show that MnSb$_2$Te$_4$ crystals have an extremely narrow hysteresis loop, i.e., they are distinctly magnetically soft materials and are comparable in coercive force to such a widely used soft ferromagnet as permalloy.  

The work was carried out within the framework of the state assignment of ISSP RAS (magnetic transport measurements), magnetic measurements were carried out with the financial support of the Russian Science Foundation (grant $\#$ 23-12-00016), and detailed data analysis was carried out with the support of St. Petersburg State University (project $\#$ 125022702939-2). Magnetic measurements were performed at the Resource Center “Center for Diagnostics of Functional Materials for Medicine, Pharmacology, and Nanoelectronics” of the St. Petersburg State University Science Park. The synthesis and characterization of samples were funded by the state budget of the Institute of Physics of the Ministry of Science and Education of Azerbaijan.

\end{document}